\documentstyle[12pt]{article}

\catcode`@=11
\def\un#1{\relax\ifmmode\@@underline#1\else
        $\@@underline{\hbox{#1}}$\relax\fi}
\catcode`@=12

\def\a{\alpha}
\def\b{\beta}
\def\c{\chi}
\def\d{\delta}

\def\g{\gamma}

\def\j{\psi}

\def\l{\lambda}

\def\p{\pi}

\def\r{\rho}
\def\s{\sigma}

\def\x{\xi}
\def\z{\zeta}

\def\L{\Lambda}

\def\ve{\varepsilon}

\def\slpa{\slash{\pa}}                            
\def\bo{{\raise-.5ex\hbox{\large$\Box$}}}               
\def\pa{\partial}                                       
\def\de{\nabla}                                         
\def\TH{{\raise.2ex\hbox{$\displaystyle \bigodot$}\mskip-4.7mu \llap H \;}}
\def\face{{\raise.2ex\hbox{$\displaystyle \bigodot$}\mskip-2.2mu \llap {$\ddot
        \smile$}}}                                      

\def\sp#1{{}^{#1}}                              
\def\sb#1{{}_{#1}}                              
\def\slash#1{\rlap{\hbox{$\mskip 1 mu /$}}#1}      
\def\Tilde#1{\widetilde{#1}}                    
\def\Bar#1{\overline{#1}}                       
\def\leftrightarrowfill{$\mathsurround=0pt \mathord\leftarrow \mkern-6mu
        \cleaders\hbox{$\mkern-2mu \mathord- \mkern-2mu$}\hfill
        \mkern-6mu \mathord\rightarrow$}
\def\dvec#1{\vbox{\ialign{##\crcr
        \leftrightarrowfill\crcr\noalign{\kern-1pt\nointerlineskip}
        $\hfil\displaystyle{#1}\hfil$\crcr}}}           
\def\dt#1{{\buildrel {\hbox{\LARGE .}} \over {#1}}}     

\def\frac#1#2{{\textstyle{#1\over\vphantom2\smash{\raise.20ex
        \hbox{$\scriptstyle{#2}$}}}}}                   
\def\sfrac#1#2{{\vphantom1\smash{\lower.5ex\hbox{\small$#1$}}\over
        \vphantom1\smash{\raise.4ex\hbox{\small$#2$}}}} 
\def\bfrac#1#2{{\vphantom1\smash{\lower.5ex\hbox{$#1$}}\over
        \vphantom1\smash{\raise.3ex\hbox{$#2$}}}}       
\def\afrac#1#2{{\vphantom1\smash{\lower.5ex\hbox{$#1$}}\over#2}}    

\def\[{\lfloor{\hskip 0.35pt}\!\!\!\lceil}
\def\]{\rfloor{\hskip 0.35pt}\!\!\!\rceil}

\def\fracm#1#2{\hbox{\large{${\frac{{#1}}{{#2}}}$}}}

\def\un{\underline}
\def\fracmm#1#2{{{#1}\over{#2}}}

\def\low#1{{\raise -3pt\hbox{${\hskip 0.75pt}\!_{#1}$}}}

\def\Dot#1{\buildrel{_{_{\hskip 0.01in}\bullet}}\over{#1}}
\def\dt#1{\Dot{#1}}

\def\Tilde#1{{\widetilde{#1}}\hskip 0.015in}

\newskip\humongous \humongous=0pt plus 1000pt minus 1000pt
\def\caja{\mathsurround=0pt}
\def\eqalign#1{\,\vcenter{\openup2\jot \caja
        \ialign{\strut \hfil$\displaystyle{##}$&$
        \displaystyle{{}##}$\hfil\crcr#1\crcr}}\,}
\newif\ifdtup

\def\ref#1{$\sp{#1)}$}

\def\pl#1#2#3{Phys.~Lett.~{\bf {#1}B} (19{#2}) #3}
\def\np#1#2#3{Nucl.~Phys.~{\bf B{#1}} (19{#2}) #3}

\def\cqg#1#2#3{Class.~and Quantum Grav.~{\bf {#1}} (19{#2}) #3}
\def\cmp#1#2#3{Commun.~Math.~Phys.~{\bf {#1}} (19{#2}) #3}

\def\mpl#1#2#3{Mod.~Phys.~Lett.~{\bf A{#1}} (19{#2}) #3}

\parskip=\medskipamount                 
\thicklines                         

\def\dslash{\not{\hbox{\kern-2pt $\partial$}}}
\def\Dslash{\not{\hbox{\kern-4pt $D$}}}
\def\pslash{\not{\hbox{\kern-2.3pt $p$}}}
 \newtoks\slashfraction
 \slashfraction={.13}
 \def\slash#1{\setbox0\hbox{$ #1 $}
 \setbox0\hbox to \the\slashfraction\wd0{\hss \box0}/\box0 }

\def\plpl{\raise-2pt\hbox{$\raise3pt\hbox{$_+$}\hskip-6.67pt\raise0.0pt}}
\def\mimi{\raise-2pt\hbox{$\raise3pt\hbox{$_-$}\hskip-6.67pt\raise0.0pt}}

\def\dvm{\raisebox{-.45ex}{\rlap{$=$}}}
\def\DM{{\scriptsize{\dvm}}~~}
\def\lin{\vrule width0.5pt height5pt depth1pt}
\def\dpx{{{ =\hskip-3.75pt{\lin}}\hskip3.75pt }}

\begin{document}


\thispagestyle{empty}               

\def\border{                                            
        \setlength{\unitlength}{1mm}
        \newcount\xco
        \newcount\yco
        \xco=-24
        \yco=12
        \begin{picture}(140,0)
        \put(-20,11){\tiny Institut f\"ur Theoretische Physik Universit\"at
Hannover~ Institut f\"ur Theoretische Physik Universit\"at Hannover~
Institut f\"ur Theoretische Physik Hannover}
        \put(-20,-241.5){\tiny University~ of~ Maryland~~ Elementary~
Particles~~
University~ of~ Maryland~~ Elementary~ Particles~~ University~ of~ Maryland~~
Elementary~ Particles~~ }
        \end{picture}
        \par\vskip-8mm}

\def\headpic{                                           
        \indent
        \setlength{\unitlength}{.8mm}
        \thinlines
        \par
        \begin{picture}(29,16)
        \put(75,16){\line(1,0){4}}
        \put(80,16){\line(1,0){4}}
        \put(85,16){\line(1,0){4}}
        \put(92,16){\line(1,0){4}}

        \put(85,0){\line(1,0){4}}
        \put(89,8){\line(1,0){3}}
        \put(92,0){\line(1,0){4}}

        \put(85,0){\line(0,1){16}}
        \put(96,0){\line(0,1){16}}
        \put(79,0){\line(0,1){16}}
        \put(80,0){\line(0,1){16}}
        \put(89,0){\line(0,1){16}}
        \put(92,0){\line(0,1){16}}
        \put(79,16){\oval(8,32)[bl]}
        \put(80,16){\oval(8,32)[br]}

        \end{picture}
        \par\vskip-6.5mm
        \thicklines}

\border\headpic {\hbox to\hsize{
\vbox{\noindent ITP--UH--15/95 \hfill April 1995 \\
UMDEPP 95-116  \hfill hep-th/9504077  }}}

\noindent
\vskip1.3cm
\begin{center}

\Large{\bf{2D $~(4,4)~$ HYPERMULTIPLETS}
\footnote{Supported in part by the `Deutsche Forschungsgemeinschaft',
\newline ${~~~~~}$ the NATO Grant CRG 930789, and the US National
Science Foundation Grant PHY-91-19746}}
\vglue.3in

{\large S. James Gates, Jr.}

{\normalsize{\it Department of Physics, University of Maryland in College Park,
\\ College Park, MD 20742-4111, USA}\\
{\sl gates@umdhep.umd.edu}} \\
\vglue.1in
{\large and \\
\vglue.1in
Sergei V. Ketov \footnote{
On leave of absence from:
High Current Electronics Institute of the Russian Academy of Sciences,
\newline ${~~~~~}$ Siberian Branch, Akademichesky~4, Tomsk 634055, Russia}}

{\normalsize{\it Institut f\"ur Theoretische Physik, Universit\"at Hannover}\\
{\it Appelstra\ss{}e 2, 30167 Hannover, Germany}\\
{\sl ketov@itp.uni-hannover.de}}
\end{center}

\vglue.2in
\begin{center}
{\Large\bf Abstract}
\end{center}

The structure of on-shell and off-shell 2D, (4,4) supersymmetric scalar
multiplets
is investigated, in components and in superspace. We reach the surprising
result
that there exist eight {\underline {distinct}} on-shell versions and an even
greater
 variety of off-shell ones.  The off-shell generalised tensor and relaxed N = 4
multiplets are introduced in superspace, and their universal invariant
self-interaction is constructed.

\newpage

{\bf 1} {\it Introduction}. It has almost always been assumed that 2D, N = 4
supersymmetry leads to relatively {\it unique} field theory representations.
Possibly for this reason, N = 4 supergravity and N = 4 superstrings have been
thought to be pretty much unique too. There has even been a proposal that an
N = 4 superstring is the paradigmatic generator of all string models \cite{bv}.
More recently, however, we have found increasing evidence that the uniqueness
may not be the case \cite{g1,k1}. We think it is useful to learn more about
manifest N = 4 supersymmetry and N = 4 scalar multiplets since these play a
crucial role in providing any Lagrangian and off-shell description of related
supergravities and superstrings. We want to maintain in any case the $SU(2)$
part of the maximal $SO(4)\cong SU(2)\times SU(2)'$ internal symmetry rotating
N = 4 supercharges. It is the $SU(2)$ that is a part of the 2D, N = 4
superconformal symmetry.
\vglue.2in

{\bf 2} {\it On-shell, 2D, N = 4 hypermultiplets}. Hypermultiplet theory
began with 4D, N = 2 models when Fayet \cite{fa} introduced this supersymmetry
representation (see ref.~\cite{so} for its description in superspace). We can
directly reduce their results to 2D, N = 4 to find the following
superdifferential equations. We call this the {\it original hypermultiplet}
(OHM) theory,
$$ D_{\a i} {\cal A}^{j \hat I} ~=~ 2 \d_i {}^j C^{\hat I \hat J}
{\Bar \psi}_{\a \hat J} ~~~ , ~~~ {\bar D}_{\a}{}^i {\cal A}^{j \hat I}
{}~=~ 2 C^{ i j} \psi_{\a} {}^{\hat I} ~~~, $$
\begin{equation}
 D_{\a i}  \psi_{\b}{}^{\hat I} ~=~ i \, C_{i j} (\g^c)_{\a \b} (\, \pa_c
{\cal A}^{j \hat I} \,) ~~~,~~~ {\bar D}_{\a}{}^i \psi_{\b}{}^{\hat I}
{}~=~ 0  ~~~.
\end{equation}
This (4,4) hypermultiplet is related to the SM-III theory in the recently
introduced classification scheme for (4,0) hypermultiplets \cite{gr}.
The proof of this can be carried out simply. In the above equations
 ${\cal A}^{i \hat I}$ is restricted to satisfy the equation
$ ({\cal A}^{i \hat I})^* = C_{i j} C_{\hat I \hat J} {\cal A}^{j \hat J}$.
This implies that not all of ${\cal A}^{i \hat I}$ is independent. A solution
to this algebraic constraint is given by ${\cal A}^{i \hat I} = ( C^{i j}
{\cal A}_j, i \,{\bar {\cal A}}^i )$ for $ \hat I = 1 ,2, $ respectively. The
equations above, rewritten in terms of ${\cal A}_i$, can then be seen
to be exactly equivalent to the SM-III theory of ref.~\cite{gr},
\newline \indent {2D, N = 4~ SM-III}
$$
D_{\a i} {\cal A}_j ~=~ C_{i j} \p_{\a} ~~,~~ {\bar D}_{\a} {}^i {\cal A}_j
{}~=~ \d_j {}^i \r_{\a} ~~,~~ {\bar D}_{\a} {}^i \r_{\b} ~=~ 0 ~~,~~
D_{\a i} \p_{\b} ~=~ 0 ~~, $$
\begin{equation}
D_{\a i} \r_{\b} ~=~ i 2  (\g^c)_{\a \b} \pa_c {\cal A}_i  ~~~; ~~~
{\bar D}_{\a} {}^i \p_{\b} ~=~ i 2 C^{i j} (\g^c)_{\a \b} \pa_c
{\cal A}_j  ~~~.
\end{equation}
However, OHM is not the only 2D, N = 4 on-shell hypermultiplet which exists.
Each of the following also forms a 2D, N = 4 on-shell representation:
\newline \indent {2D, N = 4~ SM-I}
$$ D_{\a i} {\cal A} ~=~ \varphi_{\a i} ~~~,~~~  {\bar D}_{\a}
{}^i {\bar {\cal B}} ~=~ C^{i j}  \varphi_{\a j} ~~~,~~~  {\bar D}_{\a}
{\cal A} ~=~ 0  ~~~,~~~  D_{\a i} {\Bar {\cal B}} ~=~ 0 ~~~,
$$
\begin{equation}
 D_{\a i} \varphi_{\b j} ~=~ i 2 C_{i j} (\g^c)_{\a \b} \pa_c {\Bar {\cal B}}
{}~~~,~~~ {\bar D}_{\a}{}^i \varphi_{\b j} ~=~ i 2 \d_j {}^i (\g^c)_{\a \b}
\pa_c {\cal A} ~~~;
\end{equation}
\indent {2D, N = 4~ SM-II}
$$~~~~~~{D}_{\a i } {\varphi}  ~=~
 {\l}_{\a i}  ~~~,
{}~~~{D}_{\a i } {\varphi}_j {}^k ~=~   i \left [
\d_i {}^k { \l}_{\a j} ~-~ \fracm 12 \d_j {}^k {\l}_{\a i}  \right ] ~~~,
{}~~~~~~~ ~~~~~~~~~~~~~$$
\begin{equation}
{}~~~~~~~~~~\,{D}_{\a i } {\l}_{\b j}  ~=~ 0~~~~~,~~~\,
{\bar D}_{\a}{}^{ i } {\l}_{\b j}  ~=~ i \d_j {}^i
( \g \sp{a})  \sb{\a \b}  \left ( \pa_a {\varphi} \right ) ~+~ 2
(\g \sp{a})  \sb{\a \b} \left ( \pa \sb{a} {\varphi}_{j}{}^i \right )
{}~~~.~~~ \end{equation}
\indent {2D, N = 4~ SM-IV}
\begin{equation}
 \eqalign{ {~~~~~}
{\bar D}_{\a} {}^i {\cal B}_{j}~ =~ & \d_j {}^i \, \psi_{\a} ~+~ i 2 \,
\psi_{\a} {}_j
{}^{i} ~~~~,~~~~\, {~~~~~~~~~~~~~~~~~~~~} D_{\a i} {\cal B}_{j}~ =~ 0 ~~~~,\cr
D_{\a i} \psi_{\b} ~ =~ &  i  \, (\g^c)_{\a \b} \pa_c {\cal B}_{i} ~~~~~~~~~~~,
{~~~~~~~~~~~~~~~~} {~~~~~~~~~~~~} \psi^{\a}
{}~=~ (\psi^{\a})^* ~~~~, \cr
D_{\a i} \psi_{\b} {}_j {}^{k}~ =~ &  \, \d_i {}^k (\g^c)_{\a \b} \pa_c
{\cal B}_{j} - \frac12 \d_j {}^{k} \,  (\g^c)_{\a \b} \pa_c {\cal
 B}_{i} ~~~~,~~~~ {~~~}\psi^{\a} {}_i {}^{j} ~=~ (\psi^{\a} {}_j
{}^i)^* ~~~~. }
\end{equation}

Thus, we see that the classification scheme \cite{gr} used for the (4,0)
hypermultiplets completely carries over to the case of the full (4,4)
hypermultiplets.  However, with the full (4,4) supersymmetry, there appear
even more such multiplets because we can apply parity twists to replace some
of the scalar fields in a given hypermultiplet by pseudo-scalar fields. One
such example is provided by
$$ D_{\a i} f ~=~ 2 \, C_{i j} \r_{\a}{}^j ~~~,~~~ {\bar D}_{\a}
{}^i  f ~=~  0 ~~~, ~~~ D_{\a i} g ~=~ 2 \, (\g^3)_{\a} {}^{\b} \,{\Bar
\r}_{\b i} ~~~,~~~ {\bar D}_{\a} {}^i  g ~=~  0 ~~~,
$$
\begin{equation}
 D_{\a i} \r^{\b j} ~=~ i 2 \d_i {}^j (\g^3 \g^c)_{\a} {}^{\b} (\, \pa_c
{\Bar g} \,)
{}~~~,~~~ {\bar D}_{\a}{}^i \r^{\b j} ~=~ i C^{i j}  (\g^c)_{\a} {}^{\b}
(\, \pa_c  f \,) ~~~.
\end{equation}
This particular example represents replacing two of the scalar fields
in the SM-I hypermultiplet by pseudo-scalars. A second similar example is
given by
$$
D_{\a i} {\Tilde {\cal A}} ~=~ i C_{i j} {\Tilde {\Bar \l}}_{\a} {}^j ~,~~
D_{\a i} {\Tilde {\cal B}} ~=~ - C_{i j} (\g^3)_{\a} {}^{\b}
{\Tilde {\Bar \l}}_{\b} {}^j ~,~~ D_{\a i} {\Tilde L} ~=~ i {\Tilde
{\l}}_{\a i} ~, ~~ D_{\a i} {\Tilde R} ~=~ (\g^3)_{\a} {}^{\b}
{\Tilde {\l}}_{\b i} ~,
$$
\begin{equation}
\eqalign{
D_{\a i} {\Tilde {\l}}_{\a j} &=~ - C_{i j} [~ (\g^c)_{\a \b} (\, \pa_c
{\Tilde {\cal A}}  \,) ~+~ i (\g^3 \g^c)_{\a \b} (\, \pa_c
{\Tilde {\cal B}}  \,) ~ ] ~~~, \cr
{\bar D}_{\a}{}^i {\Tilde {\l}}_{\b j} &=~ \d_j {}^i [~ (\g^c)_{\a \b} (\,
\pa_c {\Tilde L} \, ) ~+~ i (\g^3 \g^c)_{\a \b} (\, \pa_c {\Tilde R} \, ) ~]
 ~~~.  }
\end{equation}
So we see that there are many distinct on-shell hypermultiplet representations.
It is a challenge to attempt to classify how many such representations exist.
Fortunately, there is a tool available that can be used to put an upper limit
on this number. In the previous work \cite{gr}, we have been able to classify
all (4,0) hypermultiplets as well as (4,0) minus spinor multiplets. There are
four representations of each. A full on-shell (4,4) hypermultiplet is just the
sum  of a (4,0) hypermultiplet plus a (4,0) minus spinor multiplet. Therefore,
the maximum number of N = 4 hypermultiplets is sixteen. We are going to
amplify this point later on. It should be noticed that only the SM-II (RHM)
and SM-III (OHM) exist as hypermultiplet theories in 4D. The rich profusion of
 hypermultiplets is therefore a solely 2D phenomenon.
\vglue.2in

{\bf 2} {\it Off-shell hypermultiplets}. The problem of finding the off-shell
form of each on-shell hypermultiplet formulation is an unsolved one, and we are
not going to solve it in full here. \footnote{A solution may require {\it
infinite} numbers of auxiliary fields in {\it some} cases \cite{gs}, but it is
precisely \newline ${~~~~~}$ the situation we want to avoid.} The previously
known off-shell formulations of N = 4 hypermultiplets (with finite number of
auxiliary fields) include two {\it twisted} hypermultiplet versions (TM-I and
TM-II) and the `{\it relaxed' hypermultiplet} (RHM) \cite{g1}.

The twisted-I (TM-I) multiplet was the first off-shell description provided for
a 2D, N = 4 hypermultiplet. Its supersymmetry transformation laws are
$$
D_{\a i} F  ~=~ 2  C_{i j} \psi_{\a} {}^j ~~~, ~~~
D_{\a i} S  ~=~ -i  {\Bar \psi}_{\a i}  ~~~, ~~~
D_{\a i} P  ~=~  (\g^3)_{\a} {}^{\b} {\Bar \psi }_{\b i} ~~~, ~~~
$$
$$
D_{\a i} \psi^{\b j} ~=~  \d_i {}^j \left [~ (\g^c)_\a
{}^{\b} (\pa_c S) ~+~ i(\g^3 \g^c)_\a  {}^{\b} (\pa_c P)~\right ]
$$
$${~~~~~~~~~~~} + ~ \fracm 12  \left [ ~\d_i {}^j (\g^3)_{\a}
{}^{\b} A ~+~ i \d_{\a }
{}^{\b} A_i {}^j ~\right ] ~~, $$
$${~~~}
{\bar D}_{\a}{}^{i} \psi^{\b j} ~=~ i  C^{i j}
(\g^c)_\a  {}^{\b} (\pa_c F) ~~, {~~~~~~~~~~~~~} {~~~~~~~~~~~}
$$
$$
D_{\a i} A ~=~ -i 2  (\g^3 \g^c)_{\a} {}^{\b} {\pa}_c
{\Bar \psi }_{\b i}  ~~, {~~~~~~~~~~~} {~~~~~~~~~} \,
$$
\begin{equation}
D_{\a i} A_j {}^k  = {~~} 4(\d_j {}^l \d_i {}^k - \frac 12
\d_j {}^k \d_i {}^l ) (\g^c)_{\a} {}^{\b} {\pa}_c {\Bar \psi
}_{\b l}  ~~. {~~~~~} \,
\end{equation}
All the  fields are real (for $ A_i {}^j = ( A_j {}^i)^* $) with the
exception of $F$ and ${\psi}_{\a i}$ . The TM-I multiplet is
a parity twisted version of the SM-I multiplet where one scalar
field is replaced by a pseudoscalar.

The invariant component-level action takes the form
$$
{\cal S}_{{\rm T}{\rm M}{\rm {- I}} } ~=~  \int d^2 x
 ~ [~  \frac 12 S \bo S ~+~  \frac 12 P \bo P ~+~
 \frac 12 F \bo  {\Bar F} ~+~ i
{ \psi}^{\a i} ( \g \sp{c})  \sb{\a \b} \pa_c
{\Bar \psi}^{\b}{}_{i} {~~~~~~} {~~~~~~}
$$
\begin{equation}
{~~~~~~~~~~}  - \frac 12 A^2 ~-~ \frac 1{16} A_i {}^j A_j {}^i ~ ~ ] ~~~,
{~~~~~~}
\end{equation}
or in terms of unconstrained prepotentials ($V$ and $V_i {}^j$) we find
\begin{equation}
{~~~~~~~~~~}  {~~~~~~~~~~} {\cal S}_{{\rm T}{\rm M}{\rm {- I}} }
{}~=~ - \int d^2 x d^4 \z \, d^4 {\bar \z} ~[~ V  A ~+~ V_i {}^j A_j {}^i ~]
 ~~~. {~~~~~~~~~~}  {~~~~~~~~~~}
\end{equation}

The second off-shell hypermultiplet was the twisted-II (TM-II) theory
discovered by Ivanov and Krivonos \cite{ik}.  A description consistent with
their work is given by
$$~~~~~~~~~~~~~{D}_{\a i } {\cal T}  ~=~
(\g \sp{3})  \sb{\a} \sp{\b} {\Psi}_{\b i}  ~~~,
{}~~~~~~~~~~~~~~~~~~~~~~~~~~~~~~~~~~~~~~~~~~~~~~~~~~$$
$$~~~~~~~~~~~~~~~~~~~~~{D}_{\a i } {\cal X} {}_j {}^k ~=~   i \left [
\d_i {}^k { \Psi}_{\a j} ~-~ \fracm 12 \d_j {}^k {\Psi}_{\a i}  \right ] ~~~,
{}~~~~~~~~~~~~~~~~~~~~~~~~~~~~~~~~~~~~~~~~~~~~~~~~$$
$$~~~~~~~~~~~~~~~~~~~~~~~~~~~~~~ {\cal X} {}_i {}^i  ~=~ 0 ~~~, ~~~ {\cal X}
{}_i {}^j ~-~ ( {\cal X} {}_j {}^i )^* ~=~ 0 ~~~,
{}~~~~~~~~~~~~~~~~~~~~~~~~~~~~~~~~~~~~~~~~~~~~~~~~~~~$$
$$~~~~~~~~~~~~~{D}_{\a i } {\Psi}_{\b j}  ~=~ \fracm 12  C_{\a \b} C_{ i j}
{\Bar J}~~~, ~~~~~~~~~~~~~~~~~~~~~~~~~~~~~~~~~~~~~~~~~~~~~~~~~~~~~~$$
$$~~~~~~~~~~~~~~~~~~~~~~~~~~~~{D}_{\a i } {\Bar J} ~=~ 0 ~~~,~~~ m ~-~
({m} )^* ~=~ 0 ~~~, n ~-~ ({n} )^* ~=~ 0 ~~~, ~~~~~~~~~~~~~~~~~~~~~~~~~~~~$$
$${\bar D}_{\a}{}^{ i } {\Psi}_{\b j}  ~=~ i \d_j {}^i
(\g \sp{3} \g \sp{a})  \sb{\a \b}  \left ( \pa_a {\cal T} \right ) ~+~ 2
(\g \sp{a})  \sb{\a \b} \left ( \pa \sb{a} {\cal X}_{j}{}^i \right )
{}~~~~~~~$$
$$~~~~~~~~~~~~~
 ~+~ i \fracm 12 C_{\a \b} \d \sb j {}^i m ~+~ \fracm 12 (\g \sp{3})
\sb{\a \b}  \d \sb j {}^i n ~~. ~~~~~~~~~~$$
$$ ~~~~~{D}_{\a i } { J} ~=~ - i 4  C_{ i j} (\g \sp{a})  \sb{\a} \sp{\b}
\left ( \pa_a  {\Bar \Psi}_{\b}{}^{j} \right ) ~~~,~~~~~~~~~~~~~~
{}~~~~~~~~~~~ $$
$$ ~~~{D}_{\a i } n ~=~ - i 2 (\g \sp{3} \g \sp{a})  \sb{\a} \sp{\b}
\left ( \pa_a  { \Psi}_{\b i} \right ) ~~~,~~~~~~~~~~~~~~~~~~~
{}~~~~~~~~ $$
\begin{equation}
{}~{D}_{\a i } m ~=~ - 2 ( \g \sp{a})  \sb{\a} \sp{\b}
\left ( \pa_a  { \Psi}_{\b i} \right ) ~~~.~~~~~~~~~~~~~~~~
{}~~~~~~~~~~~~~
\end{equation}
Here the complex fields are $J$ and ${\Psi}_{\a i}$. This multiplet
is a parity-twisted version of the SM-II hypermultiplet, where again
one scalar is replaced by a pseudoscalar.

An invariant component-level action is
$$
{\cal S}_{{\rm T}{\rm M}{\rm {- II}}} ~=~  \int d^2
x ~ [~  \frac 12  {\cal T}  \bo  {\cal T} ~+~
 {\cal X} {}_j {}^i  \bo  {\cal X} {}_i {}^j ~+~ i
{ \Psi}^{\a}{}_{i} ( \g \sp{c})  \sb{\a \b} \pa_c
{\Bar \Psi}^{\b}{}^{i} {~~~~~~} {~~~~~~}
$$
\begin{equation}
{~~~~} {~~~~}  ~-~
\frac 18 (~ m^2 ~+~ n^2 ~+~ {J}{\Bar J} ~) ~ ] ~~.  {~~~~~~}
\end{equation}
The superfield form of this action is given by
\begin{equation}
{\cal S}_{{\rm T}{\rm M}{\rm {- II}}} ~=~ - \int d^2
x d^4 \z \, d^4 {\bar \z} ~[~ K m ~+~ L n ~]  ~-~ [~ \int d^2 x d^4 \z~
\L {J} ~+~ {\rm h}. \, {\rm c}. ~]  ~~~,
\end{equation}
in terms of the real superfield prepotentials $K$ and $L$ and chiral superfield
prepotential $\L$.

One of the interesting features of the hypermultiplet pair TM-I and TM-II is
that
they are dual to each other in such a way that they can form a supersymmetric
invariant that introduces mass without the introduction of a central charge
\cite{gr,ik}. It is the long-held but false belief that potentials for 2D
hypermultiplets require central charges \cite{af}. In terms of superfields,
this
 mass term takes the form
\begin{equation}
{~~~~~~~~~~}  {~~~~~~~~~~} {\cal S}_{{\rm N} = 4, {\rm {mass}}}
 ~=~  {M'}_0 \int d^2 x \, d^4 \z \, d^4 {\bar \z}\, \, [~ V {\cal T} ~+~
\frac 12 V_i {}^j {\cal X}_j {}^i ~]
 ~~~, {~~~~~~~~~~}  {~~~~~~~~~~}
\end{equation}
or, alternatively,
\begin{equation}
{\cal S}_{{\rm N} = 4, {\rm {mass}}} ~=~ - {\Tilde M}_0 \, \int d^2
x d^4 \z \, d^4 {\bar \z} ~ [~ K S ~+~ L P ~]   ~-~ {\Tilde M}_0
[ \, \int d^2 x d^4 \z~  \L F ~+~ {\rm h}. \, {\rm c}. ~]  ~~~.
\end{equation}
At the component level, these are equivalent to
$$
{\cal S}_{{\rm N} = 4, {\rm {mass}}} ~=~  M_0  \int d^2 \s ~[ ~
  \frac 12 m S ~-~  \frac 12 n P ~-~ \frac 14 (~ J {\Bar F} ~+~
{\Bar J} { F}~)   {~~~~~~~~~~~~~~~~~~~~~}
$$
\begin{equation}
{~~~~~~~} {~~~~~~~} {~~~~~~~~~~~~~~~} -~ \frac 18 {\cal X} {}_i {}^j
A_j {}^i ~-~ \frac 12 {\cal T} A ~+~
(~ {\Bar \Psi}^{\a i} {\Bar \psi}_{\a i} ~+~ {\rm h.} {\rm c.}
 ~) ~] ~~~.
\end{equation}

The RHM at the time of its discovery appeared as a 4D, N = 2 multiplet. It
exactly corresponds to the off-shell formulation of the SM-II theory! The
component RHM action is given by
$$
 {\cal S}_{{\rm RHM},{\rm N} = 4} ~=~  \int d^2 x ~ [~
\varphi {}_\bo \varphi  ~+~ L_{i j} {}_\bo L^{i j} ~+~ i
\psi^{\a i} \pa_{\a \b} {\Bar \psi}^{\b}{}_i {~~~~~~~~~~} {~~~~~~~~~~}
{~~~~~~~~~~~}
$$
$$ {~~~~~~~~~~~~~} +~ (~ \l^{\a i} \chi_{\a i} ~+~ {\rm h.}{\rm c.} ~)
{}~-~  2(~ \l^{\a i j k} \chi_{\a i j k} ~+~ {\rm h.}{\rm c.} ~)
$$
$$
{~~~~~~~~~~~~~~~} +~ \frac 1{18} N {\Bar N} ~+~ \frac 38 (~K^{i j}
{\Bar K}_{i j} -  M^{i j} {\Bar M}_{i j} ~) - \frac 54 C^{ i j k l}
L_{ i j k l}
$$
\begin{equation}
{~~~~~~~~~~~~~~~~~} -~ \frac 1{36} G^{\a \b} G_{\a \b} ~-~ \frac 38 (~
 A^{ \a \b i j}  A_{ \a \b i j} -  V^{\a \b i j} V_{ \a \b i j} ~) ~]
{}~~~,
\end{equation}
or, in terms of superfields, as
\begin{equation}
 {\cal S}_{{\rm RHM},{\rm N} = 4} ~=~  \int d^2 x d^4 \z d^4 {\bar \z}
{}~ [ ~ (\lambda_{\a}{}^i \rho^{\a} {}_i ~+~
{\Bar \lambda}_{\a}{}_i {\Bar \rho}^{\a} {}^i ) ~+~ L^{ijkl} X_{ijkl}
{}~] ~~~,
\end{equation}
where the unconstrained superfield potentials $ \rho^{\a} {}_i $
and $X_{ijkl}$ have been introduced.

In order to discuss the 2D, N = 4~ OHM superspace constraints, first, let us
change our notation for the superspace covariant derivatives $D_{\a}^{i\hat{I}
}$, which now carry doublet internal symmetry indices $(i,\hat{I})$, $i=1,2$,
$\hat{I}=1',2'$, of the maximal automorphism group $SO(4)\cong SU(2)\otimes
SU(2)'$ of 2D, N = 4 supersymmetry, and satisfy the reality condition
$(D_{\a}^{i\hat{I}})^* = C_{ij}C_{\hat{I}\hat{J}}D_{\a}^{j\hat{J}}$, and the
algebra $\{ D_{\a}^{i\hat{I}},D_{\b}^{j\hat{J}} \} =
iC^{ij}C^{\hat{I}\hat{J}}{\slpa}_{\a\b}$.

Being dimensionally reduced to 2D, the N = 4 OHM (= FS hypermultiplet) complex
superfields ${\cal A}_i$ satisfy the constraints
\begin{equation}
D_{\a}^{\hat{I}(i}{\cal A}^{j)}=0~~~,
\end{equation}
which put the theory on-shell, since they imply the equations of motion,
$\bo{\cal A}^i=0$. One of the ways out of this problem
\footnote{Another way is to use the harmonic superspace \cite{gikos}.}
is to introduce the {\it generalised} off-shell 2D, N = 4 tensor multiplets
$L^{i_1\cdots i_n}$, $n=2,3,\ldots$, which are defined by the constraints
\begin{equation}
D_{\a}^{\hat{I}(k}L^{i_1\cdots i_n)}=0~~~,
\end{equation}
and the reality condition $(L^{i_1\cdots i_{2p}})^*=C_{i_1j_1}\cdots
C_{i_{2p}j_{2p}}L^{i_1\cdots i_{2p}}$, in the case of an even number of
indices, $n=2p$. The tensors $L^{i_1\cdots i_n}$ are totally symmetric with
respect to their $SU(2)$ indices. In particular, when $n=2$, the superfield
$L^{ij}$ just gives the standard 4D, N = 2 tensor multiplet \cite{hst}
dimensionally reduced to 2D. \footnote{It is worthwhile noting that this
dimensional reduction is precisely equivalent to the introduction \newline
${~~~~~}$ of the TM-II multiplet.} The generalised 4D, N = 2 tensor multiplets
were introduced in ref.~\cite{ks}. The off-shell 2D, N = 4 generalised tensor
multiplets can be used to `relax' the constraints for the OHM $(n=1)$ and the
ordinary tensor multiplet $(n=2)$, for example
\begin{equation}
D_{\a}^{\hat{I}(i}{\cal A}^{j)}=D_{\a k}^{\hat{I}}{\cal A}^{ijk}~,\qquad
D_{\a}^{\hat{I}(i}{\cal A}^{jkl)}=0~~~,
\end{equation}
or
\begin{equation}
D_{\a}^{\hat{I}(i}L^{jk)}=D_{\a l}^{\hat{I}}L^{ijkl}~,\qquad
D_{\a}^{\hat{I}(i}L^{jklm)}=0~~~,
\end{equation}
which define the relaxed multiplets of the type (1--3) and (2--4),
respectively, according to the number of the external $SU(2)$ indices involved.
More general constructions of the type (1--3--$\ldots$--(2q+1)) or
(2--4--$\ldots$--(2q)) can also be introduced \cite{ks}. In particular, the
case of (2--4) defines the relaxed hypermultiplet of ref.~\cite{hst}. The
system of tensor superfields with infinite relaxation $(q=\infty)$
{\it precisely} corresponds to the harmonic superfields of ref.~\cite{gikos},
where these tensor superfields appear as the coefficients at harmonic zweibein
monomials. All such constructions are just different off-shell realizations of
 N = 4 hypermultiplet, with finite numbers of auxiliary fields.

One of the interesting tools that worked well as a way to provide
a uniform classification of (4,0) hypermultiplets \footnote{For a
earlier and different view of these theories see ref.~\cite{EvOv}.}
was the use of `spectroscopic analysis' as a way to describe all of the (4,0)
hypermultiplets \cite{gr}. A simple extension of that works for the
(4,4) case too. The four basic hypermultiplets listed above in eqs.~(1)--(5)
 can be thought of as
\begin{center}
\renewcommand\arraystretch{1.2}
\begin{tabular}{|c|c|c| }\hline
${~~~}{\rm (4,4)~ HM}{~~~}$  & ${\rm Spin}$-${\rm 0~SU(2)~Rep}^{{\rm Parity}}$
& ${\rm Spin}$-${\fracm 12} {\rm~SU(2)~Rep}^{{\rm Parity}}$  \\ \hline
\hline
${\rm SM-I}$ & ${~~~~4s^+~~~~}$ &  ${~~\fracm 12 ~~~}$ \\ \hline
${\rm SM-II}$ & ${~~1s^+1p^+~~~~}$ &  ${~~\fracm 12 ~~~}$ \\ \hline
${\rm SM-III}$ & ${~~\fracm 12 ~~~}$ &   ${~~~~4s^+~~~~}$ \\ \hline
${\rm SM-IV}$ &  ${~~\fracm 12 ~~~}$ &  ${~~1s^+1p^+~~~~}$  \\ \hline
\end{tabular}
\end{center}
\vskip.1in
\centerline{{\bf Table I}}
where we use a notation with a $+$ superscript for a scalar spin-0 field (or
a spinor) and a $-$ superscript for a pseudoscalar spin-0 field (or an
axial spinor).  It should be noticed that the definition of parity requires
spinors of both $+$ and $-$ type to be in the supermultiplet. Since for the
heterotic case only one handedness was present, there was no need to introduce
this degree of freedom in the classification scheme.

It is now clear how we should think of the additional hypermultiplets in
eqs.~(6), (7), (8) and (11). These are just the cases of spin-0 combinations
 $2s^+ 2s^-$, $3s^+ s^-$, and $1s^- 1p^+$, respectively.  A complete
enumeration of all independent \footnote{We regard a multiplet and one of its
twisted versions to be the same if one can be obtained from \newline ${~~~~~}$
the other by a simple redefinition involving $\g^3$ acting on the spinor in
the supermultiplet.  If \newline ${~~~~~}$ this is {\it {not}} the case we
say the two multiplets are {\it {independent}}.} multiplets consists of the
spin-0 combinations
\begin{equation}
 4s^+ ~~, ~~ 3s^+ s^- ~~,~~ 2s^+ 2s^- ~~,~~ 1s^+1p^+ ~~,~~ 1s^+1p^-~,
\end{equation}
as well as the spin-1/2 combinations
\begin{equation}
 4s^+ ~~, ~~ 1s^+1p^+ ~~.
\end{equation}
The spectroscopic analysis suggests that there are seven 2D independent
hypermultiplets. However, there is actaully a two-fold degeneracy in the
$2s^+ 2s^-$ case (see equations 6 and 7). So this ultimately gives eight
multiplets. \vglue.2in

{\bf 4} {\it (4,0) analysis of on-shell N = 4 hypermultiplets}. The scalars and
spinors of all the (4,0) hypermultiplets actually form real spinor
representations of $Spin(2,2)$ \cite{gr}. The same statement is true for the
minus spinor multiplets (heterotic fermion multiplets) too.

The (4,0) hypermultiplets SM-I and SM-II can be described in terms of four real
spin-0 fields denoted by $\varphi_A$ and four Majorana spinors
denoted by ${\Psi}^{-} {}_{\hat A}$ whose supersymmetry variations take the
form
\begin{equation} \d_Q \varphi_A ~=~ i \a^{+ ~ p} ( {\rm L}_p )_A {}^{\hat A}
{\Psi}^- {}_{\hat A}   ~~~, ~~~ \d_Q {\Psi}^{-} {}_{\hat A}
 ~=~   \a^{+ ~ p} ( {\rm R}_p )_{\hat A} {}^A   \pa_{\dpx}
 \varphi_A~,
\end{equation}
in terms of four real constant Grassmann parameters $\a^{+ ~p}$.
The real quantities $(
{\rm L}_p )_A {}^{\hat A}$ and $ ( {\rm R}_p )_{\hat A} {}^A$ satisfy
\begin{equation}
( {\rm L}_p )_A {}^{\hat A} ( {\rm R}_q )_{\hat A} {}^B ~+~ ( {\rm L}_q )_A
{}^{\hat A}
( {\rm R}_p )_{\hat A} {}^B ~=~ - 2 \d_{p  q} ~ ({\rm I})_A {}^B ~~~,
\end{equation}
\begin{equation}
( {\rm R}_p )_{\hat A} {}^A ( {\rm L}_q )_A {}^{\hat B} ~+~ ( {\rm R}_q
)_{\hat A} {}^A
( {\rm L}_p )_A {}^{\hat B}  ~=~ - 2 \d_{p  q} ~ ({\rm I})_{\hat A} {}^{\hat B}
{}~~~,
\end{equation}
and the ${\rm L}$-matrices and ${\rm R}$-matrices are thus generalised
$4\times 4$ Pauli matrices.  The SM-I multiplet is associated
with the set
\begin{equation}
 \eqalign{
{\rm L}_1 &=~ i \s^1 \otimes \s^2 ~~;~~{\rm L}_2 ~=~ i \s^2 \otimes {\rm I} ~~;
 ~~{\rm L}_3 ~=~ - i \s^3 \otimes \s^2 ~~;~~{\rm L}_4 ~=~ - {\rm I}
\otimes {\rm I} ~~; \cr
{\rm R}_1 &=~ i \s^1 \otimes \s^2 ~~;~~ {\rm R}_2 ~=~ i \s^2
\otimes {\rm I} ~~ ; ~~ {\rm R}_3 ~=~ - i \s^3 \otimes \s^2 ~~;~~
{\rm R}_4 ~=~ +\, {\rm I} \otimes {\rm I} ~~, }
\end{equation}
and the SM-II multiplet is associated with
\begin{equation}
 \eqalign{
{\rm L}_1 &=~ i \s^2 \otimes \s^3 ~~; ~~{\rm L}_2 ~=~ - i {\rm I} \otimes
\s^2 ~~;~~ {\rm L}_3 ~=~ i \s^2 \otimes \s^1 ~~ ; ~~ {\rm L}_4 ~=~+\, {\rm I}
\otimes {\rm I} ~~; \cr {\rm R}_1 &=~ i \s^2 \otimes \s^3 ~~;~~ {\rm R}_2 ~=~
- i {\rm I} \otimes \s^2 ~~ ; ~~{\rm R}_3 ~=~ i \s^2 \otimes \s^1 ~~; ~~
{\rm R}_4 ~=~ -  {\rm I} \otimes {\rm I} ~~. } \end{equation}
For the SM-III and SM-IV multiplets, the four real scalar fields are denoted
by $\varphi_{\hat A}$ and the four real spinors by ${\Psi}^- {}_A$ with
supersymmetry variations
\begin{equation}
 \d_Q \varphi_{\hat A} ~=~ i  \a^{+ ~ p} ( {\rm R}_p )_{\hat A} {}^A
{\Psi}^- {}_A   ~~~, ~~~ \d_Q {\Psi}^- {}_A
 ~=~   \a^{+ ~ p} ( {\rm L}_p )_A {}^{\hat A} \pa_{\dpx} \varphi_{\hat A}
{}~~~.
\end{equation}
The SM-III multiplet is associated with the set in eq.~(28) and SM-IV multiplet
is associated with the set in eq.~(29).

Very similar results follow for the spinor multiplets. In real notation, MSM-I
and MSM-II take the respective forms (below $ {\rm F} {}_{\hat A}$
denote the auxiliary fields)
\begin{equation}
 \d_Q {\Psi}^+{}_A ~=~ i  \a^{+ ~ p} ( {\rm L}_p )_A {}^{\hat A}
{\rm F} {}_{\hat A}   ~~~, ~~~ \d_Q {\rm F} {}_{\hat A}
 ~=~   \a^{+ ~ p} ( {\rm R}_p )_{\hat A} {}^A \pa_{\dpx} {\Psi}^+{}_A ~~~,
\end{equation}
where MSM-I is associated with the representation in eq.~(28) and MSM-II is
associated with the representation in eq.~(29). For MSM-III and MSM-IV
we have
\begin{equation}
 \d_Q {\Psi}^+{}_{\hat A} ~=~ i  \a^{+ ~ p} ( {\rm R}_p )_{\hat A} {}^A
{\rm F} {}_A   ~~~, ~~~ \d_Q {\rm F} {}_A ~=~   \a^{+ ~ p} ( {\rm L}_p )_A
{}^{\hat A} \pa_{\dpx} {\Psi}^+{}_{\hat A} ~~~,
\end{equation}
where MSM-III is associated with the representation in eq.~(28) and
MSM-IV is associated with the representation in eq.~(29).

Our task now is to investigate how many ways we can glue the (4,0) spinor
multiplets to the (4,0) hypermultiplets to obtain an on-shell (4,4)
supersymmetry representation. Since we are only considering on-shell
theories, we set the auxiliary fields to zero. Also any time the
Dirac equation appears, it can be set to zero.

If we attempt to `glue' the (4,0) SM-I or SM-II multiplets to either
MSM-I or MSM-II, the form of the supersymmetry variations can only be
$$\d_Q \varphi_A ~=~ i \a^{+ ~ p} ( {\rm L}_p )_A {}^{\hat A}
{\Psi}^- {}_{\hat A} ~+~ i \b^{- ~ p} ( {\rm J}_p )_A {}^{B}
{\Psi}^+ {}_{B} ~~~,  $$
\begin{equation}
\d_Q {\Psi}^{-} {}_{\hat A} ~=~   \a^{+ ~ p} ( {\rm R}_p )_{\hat A} {}^A
\pa_{\dpx} \varphi_A ~~~, \quad
\d_Q {\Psi}^{+} {}_{A} ~=~   \b^{- ~ p} ( {\Tilde {\rm J}}_p )_{A} {}^B
\pa_{\DM} \varphi_B  ~~~.
\end{equation}
In the attempt to `glue' the (4,0) SM-I or SM-II multiplets to either
MSM-III or MSM-IV, the form of the supersymmetry variations can only be
$$\d_Q \varphi_A ~=~ i \a^{+ ~ p} ( {\rm L}_p )_A {}^{\hat A}
{\Psi}^- {}_{\hat A} ~+~ i \b^{- ~ p} ( {\rm K}_p )_A {}^{\hat B}
{\Psi}^+ {}_{\hat B} ~~~,  $$
\begin{equation}
\d_Q {\Psi}^{-} {}_{\hat A} ~=~   \a^{+ ~ p} ( {\rm R}_p )_{\hat A} {}^A
\pa_{\dpx} \varphi_A ~~~, \quad
\d_Q {\Psi}^{+} {}_{\hat A} ~=~   \b^{- ~ p} ( {\Tilde {\rm K}}_p )_{\hat A}
{}^B  \pa_{\DM} \varphi_B  ~~~.
\end{equation}

The attempt to extend SM-III and SM-IV to full on-shell theories means
that (4,4) supersymmetry variations must take the forms
$$ \d_Q \varphi_{\hat A} ~=~ i  \a^{+ ~ p} ( {\rm R}_p )_{\hat A} {}^A
{\Psi}^- {}_A   ~+~ i \b^{- ~ p} ( {\cal P}_p )_{\hat A} {}^{B}
{\Psi}^+ {}_{B} ~~~,
$$
\begin{equation}
\d_Q {\Psi}^- {}_A  ~=~   \a^{+ ~ p} ( {\rm L}_p )_A {}^{\hat A}
\pa_{\dpx} \varphi_{\hat A} ~~~, \quad
\d_Q {\Psi}^{+} {}_{A} ~=~   \b^{- ~ p} ( {\Tilde {\cal P}}_p )_{A}
{}^{\hat B}  \pa_{\DM} \varphi_{\hat B}  ~~~,
 \end{equation}
when `gluing' to either MSM-I or MSM-II multiplets. Similarly, the extension
of SM-III and SM-IV to full on-shell (4,4) theories means
that (4,4) supersymmetry variations must take the forms
$$ \d_Q \varphi_{\hat A} ~=~ i  \a^{+ ~ p} ( {\rm R}_p )_{\hat A} {}^A
{\Psi}^- {}_A   ~+~ i \b^{- ~ p} ( {\cal Q}_p )_{\hat A} {}^{\hat B}
{\Psi}^+ {}_{\hat B} ~~~,$$
\begin{equation}
\d_Q {\Psi}^- {}_A  ~=~   \a^{+ ~ p} ( {\rm L}_p )_A {}^{\hat A}
\pa_{\dpx} \varphi_{\hat A} ~~~, \quad
\d_Q {\Psi}^{+} {}_{\hat A} ~=~   \b^{- ~ p} ( {\Tilde {\cal Q}}_p )_{\hat A}
{}^{\hat B}  \pa_{\DM} \varphi_{\hat B}  ~~~,
 \end{equation}
when `gluing' to either MSM-III or MSM-IV multiplets. The condition
for full on-shell (4,4) supersymmetry is precisely that the operator equation
\begin{equation}
 [ \d_Q(1) ~,~ \d_Q(2) ] ~=~ i 2 \, \d_{p q} ( \a^{+ p}_1 \a^{+ q}_2 \pa_{\dpx}
+ \b^{- p}_1 \b^{- q}_2 \pa_{\DM} ) ~~~ {~~~}
\end{equation}
is satisfied on all fields subject to the use of the Dirac equation on spinors.
This will be satisfied if
$$
( {\rm J}_p )_A {}^{B} ( {\Tilde {\rm J}}_q )_{B} {}^C ~+~ ( {\rm J}_q )_A
{}^{B}
( {\Tilde {\rm J}}_p )_{B} {}^C ~=~ - 2 \d_{p  q} ~ ({\rm I})_B {}^C ~~~,
\eqno(a)$$
$$
({\Tilde{\rm J}}_p )_A {}^{B} ({\rm J}_q )_{B} {}^C ~+~ ( {\Tilde{\rm J}}_q )_A
 {}^{B}
( {\rm J}_p )_{B} {}^C ~=~ - 2 \d_{p  q} ~ ({\rm I})_B {}^C ~~~, \eqno(b)$$
$$
( {\rm K}_p )_A {}^{\hat A} ( {\Tilde {\rm K}}_q )_{\hat A} {}^B ~+~ (
{\rm K}_q )_A {}^{\hat A}
( {\Tilde {\rm K}}_p )_{\hat A} {}^B ~=~ - 2 \d_{p  q} ~ ({\rm I})_A {}^B ~~~,
\eqno(c)$$
$$
( {\Tilde {\rm K}}_p )_{\hat A} {}^A ( {\rm K}_q )_A {}^{\hat B} ~+~
( {\Tilde {\rm K}}_q )_{\hat A} {}^A
( {\rm K}_p )_A {}^{\hat B}  ~=~ - 2 \d_{p  q} ~ ({\rm I})_{\hat A}
{}^{\hat B} ~~~,\eqno(d)$$
$$
( {\Tilde {\cal P}}_p )_A {}^{\hat A} ( {\cal P}_q )_{\hat A} {}^B ~+~ (
{\Tilde {\cal P}}_q )_A {}^{\hat A}
( {\cal P}_p )_{\hat A} {}^B ~=~ - 2 \d_{p  q} ~ ({\rm I})_A {}^B ~~~,
\eqno(e)$$
$$
( {\cal P}_p )_{\hat A} {}^A ( {\Tilde {\cal P}}_q )_A {}^{\hat B} ~+~
( {\cal P}_q )_{\hat A} {}^A
( {\Tilde {\cal P}}_p )_A {}^{\hat B}  ~=~ - 2 \d_{p  q} ~ ({\rm I})_{\hat A}
{}^{\hat B} ~~~,\eqno(f)$$
$$
({\cal Q}_p )_{\hat A} {}^{\hat B} ( {\Tilde {\cal Q}}_q )_{\hat B} {}^{\hat C}
 ~+~ ( {\cal Q}_q )_{\hat A} {}^{\hat B} ( {\Tilde {\cal Q}}_p )_{\hat B}
{}^{\hat C} ~=~ - 2 \d_{p  q} ~ ({\rm I})_{\hat B} {}^{\hat C}
{}~~~,\eqno(g)$$
$$
( {\Tilde{\cal Q}}_p )_{\hat A} {}^{\hat B} ({\cal Q}_q )_{\hat B} {}^{\hat C}
 ~+~ ( {\Tilde{\cal Q}}_q )_{\hat A} {}^{\hat B} ( {\cal Q}_p )_{\hat B}
{}^{\hat C} ~=~ - 2 \d_{p  q} ~ ({\rm I})_{\hat B} {}^{\hat C} ~~~, \eqno(h)$$
with {\underline{no}} other restrictions required! It is a fact that there
are no set of four independent tensors (with the appropriate index structure)
that satisfy equations $a,b,g$ and $h$. \footnote{Interestingly enough, these
equations do have solutions for (4,3) hypermultiplets!} We thus conclude that
there can be only eight on-shell 2D hypermultiplets, in agreement with the
spectroscopic analysis.
\vglue.2in

{\bf 5} {\it (4,4) hypermultiplet NLSM}. Remarkably, there exists the universal
N = 4 supersymmetric {\it non-linear sigma-model} (NLSM) action for any kind
(and number) of the generalised and/or relaxed tensor multiplets. First, let us
introduce the function $G(L^{i_1\cdots i_n})$ as a solution to the equations
\begin{equation}
\de_{\a}^{\hat{I}}G \equiv \left( D_{\a}^{\hat{I}1} +\x D_{\a}^{\hat{I}2}
\right) G = 0 ~~~,
\end{equation}
where a complex projective parameter $\x$ has been introduced. It is not
difficult to check that the general solution to eq.~(38) can be represented
in the form ({\it cf.\/} ref.~\cite{ks})
\begin{equation}
G=G\left( \x, Q_n(\x)\right)~,\qquad Q_n(\x)\equiv \x_{i_1}\cdots\x_{i_n}
L^{i_1\cdots i_n}~,\qquad \x_i\equiv(1,\x)~~~,
\end{equation}
where the function $G$ on the r.h.s. of this equation is now an {\it arbitrary}
differentiable meromorphic function of $\x$ and $Q_n$'s. In the case of the
relaxed hypermultiplets (21) and (22), one should use
\begin{equation}
Q_{1{\rm R}}(\x)=Q_1(\x) -\fracmm{4}{3}\fracmm{\pa Q_3}{\pa\x}~,\qquad
Q_{2{\rm R}}(\x)=Q_2(\x) -\fracmm{5}{4}\fracmm{\pa Q_4}{\pa\x}~,
\end{equation}
instead of $Q_1$ and $Q_2$, respectively, while any dependence on $Q_3(\x)$ or
$Q_4(\x)$ is also allowed. The function $G\left( \x, Q_n(\x)\right)$ is chiral
in the sense of eq.~(38). Therefore, integrating it over the remaining
superspace coordinates results in the invariant action ({\it cf.\/}
refs.~\cite{klr,lir})
\begin{equation}
S_{\rm NLSM} = \int d^2x\, \fracmm{1}{2\p i}\oint_C \fracmm{d\x}{(1+\x^2)^4}
C_{\hat{I}\hat{J}}C^{\a\b}\tilde{\de}^{\hat{I}}_{\a}\tilde{\de}^{\hat{J}}_{\b}
G\left( \x, Q_n(\x)\right) + {\rm h.c.}~~~,
\end{equation}
where the new, linearly independent on $\de$'s, superspace derivatives
\begin{equation}
\tilde{\de}^{\hat{I}}_{\a} = \x D^{\hat{I}1}_{\a} -D^{\hat{I}2}_{\a}~~~,
\end{equation}
have been introduced. The contour $C$ in the complex $\x$-plane should be
chosen in such a way that the points $\x_{\rm c}=\pm i$, where the linear
independence of $\de$'s and $\tilde{\de}$'s breaks down, will be outside the
contour.

This construction of invariant NLSM action aparently breaks down one of the
$SU(2)$ internal symmetries, but maintains another $SU(2)'$, which is just
necessary for the full 2D, N = 4 superconformal symmetry to be ultimately
represented by the `small\/' linear N = 4 superconformal algebra, from the
viewpoint of conformal field theory \cite{book}. Still, there is a chance of
having the full $SO(4)$ internal symmetry (and, hence, a larger N = 4
superconformal algebra in the corresponding N = 4 superconformal field theory),
when the function $G$ and the contour $C$ are specially chosen. Indeed, $\x$ is
the inhomogeneous $CP(1)$ coordinate, whose $SU(2)$ transformation law is given
by
\begin{equation}
\x' = \fracmm{\bar{a}\x-\bar{b}}{a+b\x}~,\qquad \left(\begin{array}{cc}
a & b \\ -\bar{b} & \bar{a} \end{array}\right)\in SU(2)~,
\qquad a\bar{a}+b\bar{b}= 1~~~.
\end{equation}
This obviously implies $Q_n'(\x')=(a+b\x)^{-n}Q_n(\x)$. Hence, the action (41)
will be $SO(4)$ invariant provided that
\begin{equation}
G(\x',Q_n')=(a+b\x)^{-2}G(\x,Q_n)~~~,
\end{equation}
up to an additive total derivative.

It is presently believed \cite{so} that the OHM does not allow any non-trivial
off-shell formulation with a finite number of auxiliary fields. This is,
however, not in conflict with our results, since (i) any theory in terms of
the relaxed off-shell combination $Q_{1{\rm R}}$ actually has a larger number
of propagating degrees of freedom, and (ii) there is {\underline{no}} on-shell
condition in the case of the generalised tensor (or relaxed tensor) multiplets
$(n\geq 2)$, unlike the OHM case of $n=1$.

Among the components of the 4D, N = 2 generalised tensor multiplet,
\begin{equation}
L^{i_1\cdots i_n}, \quad \j_{\a}^{i_1\cdots i_{n-1}},\quad
C^{i_1\cdots i_{n-2}}, \quad V_{\a\dt{\a}}^{i_1\cdots i_{n-2}}, \quad
\c_{\a}^{i_1\cdots i_{n-3}}, \quad D^{i_1\cdots i_{n-4}}~~~,
\end{equation}
there is a 4D vector $V$, which is only conserved when $n=2$. \footnote{It can
 be easily checked by counting the numbers of the off-shell bosonic and
fermionic
degrees ${~~~~~}$ of freedom which must coincide.}  The vector fields for
$n>2$ can be easily eliminated via their algebraic equations of motion in the
NLSM action, whereas in the case of $n=2$ the dimensional reduction of the 4D
conserved vector results in the 2D conserved vector ${V'}_a$ and two auxiliary
scalars. The latter also have algebraic equations of motion, whereas the
former can be substituted by a propagating scalar $B$ via
${V'}_a=\ve_a{}^b\pa_bB$, which results in the NLSM torsion. The situation is
similar in the case of the 4D, N = 2 vector multiplets dimensionally reduced
to 2D \cite{krev}. Therefore, it is the presence of the TM-I and TM-II
multiplets that introduces torsion in the 2D, N = 4 NLSM. \footnote{This gives
another reason to call them `twisted'.} Unlike the 4D, N= 2 NLSM \cite{ks,klr}
 which is non-renormalisable and does not always have simple geometrical
interpretation, the 2D, N=4~ NLSM of eq.~(41) is either hyper-K\"ahlerian
(in the absence of torsion) or, at least, quaternionic, and it is UV finite to
 all orders of perturbation theory, besause of (4,4) supersymmetry \cite{krev}.

The existence of many distinct 2D hypermultiplets implies the existence of many
distinct N = 4 `{\it mirror maps}' between them, as well as between the
corresponding NLSM's. They are the N = 4 analogues of the `mirror symmetry'
familiar from the N = 2 case.

{\it Note added.} After our paper was completed, we have been informed that the
harmonic superspace description of the interacting TM-II had been recently
given by
E. Ivanov and A. Sutulin in Nucl.~Phys.~{\bf B432}~(1994)~246.
\vglue.2in

\noindent {\it Acknowledgement}:
One of the authors (SVK) acknowledges useful discussions with Emery Sokatchev.

\end{document}